\documentclass[12pt]{article}
\usepackage{graphicx}
\begin{document}
\title{
\hfill {\rm\small MCTP-02-14} \\
\vskip-15pt~\\ A Wavelet Analysis of Solar Climate Forcing: \\ I) Solar
Cycle Timescales}

\author{Matthew J. Lewis and Katherine Freese\\ \\ {\rm\small {\it Michigan
Center for Theoretical Physics, University of Michigan, Ann Arbor, MI
48109, USA}}}

\date{}
\maketitle

\begin{abstract}

We use the technique of wavelet analysis to quantitatively investigate
the role of solar variability in forcing terrestrial climate change on
solar cycle timescales (roughly 11 years).  We examine the connection
between mean annual solar irradiance, as reconstructed from sunspot
and isotope records, and the climate, as proxied by two Northern
Hemisphere surface air temperature reconstructions. By applying
wavelet transforms to these signals, we are able to analyze both the
frequency content of each signal, and also the time dependence of that
content. After computing wavelet transforms of the data, we perform
correlation analyses on the wavelet transforms of irradiance and
temperature via two techniques: the Pearson's method and the
conditional probability method. We thus track the correlation between
individual frequency components of both signals as a function of
time. A nonzero correlation between the irradiance and temperature
wavelet spectra requires a phase lag between the two data sets ({\it
i.e.}, terrestrial response to solar output is not instantaneous). We
search for the optimal phase lag that maximizes the correlation. By
choosing an appropriate phase for each year, we find a significant,
positive sun-climate correlation for most of the period AD
1720-1950. We find that this phase-optimized correlation varies in
time, oscillating between 0.12 and 0.71 throughout the past 400
years. We find that the phase lag varies from 0-10 years. We also
present a test to determine whether terrestrial response to solar
output is enhanced via stochastic resonance, wherein the weak periodic
solar signal is amplified by terrestrial noise.

\end{abstract}

\section{Introduction}

{

Much work in recent years has been directed at assessing the role of
the sun in influencing climate change, but many questions remain. On
the timescale of the roughly 11-year solar cycle, causal connections
between the sun and climate almost certainly exist but have been
overinterpreted in the literature without firm quantitative study: the
observed correlations are only apparent in select records and are
highly transient, and the mechanisms behind the large response of
terrestrial temperature to small fluctuations in solar output remain a
puzzle [Pittock, 1978; Burroughs, 1992]. Understanding the
sun-climate connection on all timescales is particularly important
today because it is crucial to assessing anthropogenic climate impact.

We examine the connection between mean annual solar irradiance, as
reconstructed from sunspot and isotope records, and the climate, as
proxied by Northern Hemisphere surface air temperature
reconstructions. We use the total solar irradiance reconstruction of
Lean, Beer, and Bradley [1995], which extends back to AD 1610. As our
indicators of climate, we use the multi-proxy surface air temperature
(SAT) reconstructions of Mann, Bradley, and Hughes [1998] as well as
those of Jones, {\it et al.} [1998]. The solar irradiance and SAT
reconstructions of Lean, Beer, and Bradley [1995] and Mann, Bradley,
and Hughes [1998] are plotted in Figure \ref{mannleancomparison} for
comparison. From an examination of the two data sets by eye, one might
naively conclude that the sun and temperature track one another so
closely that the recent high terrestrial temperatures are caused by
increases in solar irradiance. It is our goal to instead perform a
quantitative study of the sun-earth connection using wavelet analysis.
As a first step, in this paper we study the sun-earth connection on
the roughly 11-year timescale of the solar cycle. A future paper will
study the sun-earth connection on century timescales, with the goal of
separating the importance of solar and greenhouse contributions to
recent highs in earth temperatures.

\begin{figure}[ht]
\begin{center}
\includegraphics[scale=1]{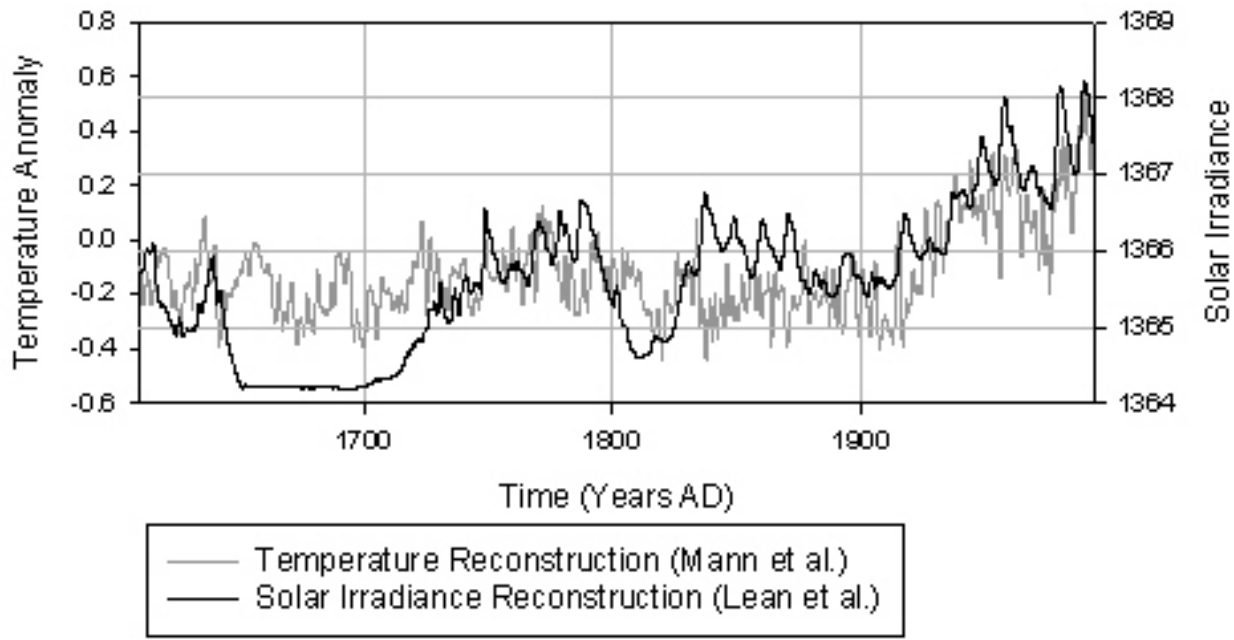}
\end{center}
\caption{ A comparison of the solar irradiance and surface air
temperature reconstructions.}
\label{mannleancomparison}
\end{figure}

We approach the problem of quantifying the sun-climate connection on
decadal timescales by examining the correlation between spectral
features of solar irradiance and global temperature reconstructions as
a function of time. Using the method of wavelet analysis, we can
objectively determine the dominant solar forcing frequencies in any
given year. We can quantitatively compare the observed period and
phase with the corresponding spectral characteristics of the
temperature reconstruction. In effect, we can examine the sun-climate
correlation as a function of both time and frequency.

While correlation studies between solar irradiance and climate
variables have been done before [Crowley and Kim, 1996;
Friis-Christensen and Lassen, 1991; Hoyt and Schatten, 1997; McCormack
and Hood, 1996], this study offers several new benefits. Wavelets
allow us to determine, in any given year, what frequencies present in
the solar reconstruction are also present in the temperature data. For
example, in 1976 we detect the presence of an 11.54 year solar cycle
in the irradiance record; we find the same cycle evident in the
temperature record with a phase lag of a few years.

Determining the correlation between the spectral content of two
signals offers advantages over comparing directly the raw time
signals: the wavelet transform acts as a filter, allowing us to
examine frequencies of interest while excluding much of the stochastic
variability and background noise inherent in the original
signal. Furthermore, reliable techniques exist for establishing the
statistical validity of the results [Torrence and Compo, 1998].  By
applying wavelet transforms to these signals, we are able to analyze
both the frequency content of each signal, and also the time
dependence of that content.

We compute wavelet transforms of the solar and temperature
reconstructions, and compare them using two different schemes: the
Pearson's correlation method and the conditional probability method.
In this paper we focus on timescales of the (approximately) 11-year
solar cycle.  The results of our analysis show that a nonzero
correlation between solar irradiance and terrestrial temperature on
this timescale requires a phase lag between the two data sets; this is
not surprising since we expect terrestrial response to lag behind
solar output. Hence we study the correlation between wavelet
transforms of solar output in one year with wavelet transforms of
terrestrial temperature an arbitrary number of years later. We search
for the optimal phase lag that maximizes the correlation. By choosing
an appropriate phase shift for each year, we do find a significant,
positive sun-climate correlation for most of the period AD 1720-1950.

Because the solar cycle length varies in time, we improve our results
further by identifying the appropriate solar-cycle length as a
function of time, again using wavelets. Then we compute the
irradiance-temperature wavelet spectra correlation at the corrected
timescale. We find that the phase-optimized correlations oscillate in
time. In the case of the sun-climate correlation corrected for phase
lag and solar cycle length, we find that the strength of the
correlation is not at all constant in time: in fact, the correlation
increases and decreases between 0.12 and 0.71 (in a range between -1
and 1) over several intervals throughout the past 400 years. We find
that the phase lag varies from 0-10 years.

We present also a test for the possible role of stochastic resonance
in the sun-climate system. Stochastic resonance is a mechanism whereby
a weak, periodic signal is amplified by the noise associated with a
nonlinear system with more than one minimum. We suggest here a test to
determine whether terrestrial response to solar output is enhanced via
stochastic resonance.

Many authors have addressed the issue of solar climate forcing,
including: [Andronova and Schlesinger, 2000; Reid, 2000; van Loon and
Labitzke, 2000; Rind, Lean, and Healy, 1999; Karl and Trenberth,1999;
Mann, Bradley, and Hughes, 1998; North and Stevens, 1998; White, {\it
et al.}, 1997]. General circulation models (GCMs) provide an
alternative approach to investigate solar climate forcing by
simulating climate dynamics [Andronova and Schlesinger, 2000; Rind,
Lean, and Healy, 1999; Marshall, {\it et al.} 1994]. This paper
presents work complementary to such efforts by investigating directly
the relationship between reconstructed sun and climate data sets.

In Section II we discuss the paleoclimate reconstructions that we use.
In Section III we review the basic theory of wavelet analysis, and
then obtain wavelet transforms of both the solar irradiance and the
surface air temperature data sets. In Section IV, we use two different
methods to obtain wavelet correlations between the solar irradiance
and terrestrial temperature data sets, in order to compare the 11-year
wavelet transforms. We use the Pearson's method as well as the
conditional probabilities method. In Section V we discuss the results
of our study. We introduce in Section VI a proposal for using wavelets
to extract information about the existence of stochastic resonance in
the data. We conclude in section VII with a summary of our results.

This paper deals with the sun-climate connection on decadal
timescales, focusing on the approximately 11-year solar cycle. Future
work will investigate the solar-climate connection on longer
timescales of centuries to millenia, with the goal of addressing the
solar vs. anthropogenic contributions to increasing terrestrial
temperatures.

\section{Paleoclimate Reconstructions}

Since we are interested in assessing the sun-climate connection on
decadal time scales, we need robust solar and climate indicators
extending back hundreds of years. Reliable global temperature records
exist for only the past century, and measurements of solar irradiance
have only been made for twenty years. To work beyond the limitations
of instrumental records, we must turn to paleoclimate reconstructions.

\subsection{Solar Activity}

The intensity of radiation from the sun reaching the surface of the
earth, currently about $1365$ W/m$^2$, is subject to cyclic behavior
on several time scales. The most prominent of these variations are the
well known 11-year solar cycle and its 22-year sub-harmonic or Hale
cycle, and evidence also exists for longer term variability,
particularly the 88-year Gleissberg cycle [Hoyt and Schatten, 1997;
Schatten, 1988]. In addition, millennial-scale solar variability is
present as a result of changes in the earth's orbit [Berger, 1991].
As a measure of decadal and centennial-scale solar variability, we use
the total solar irradiance reconstruction of Lean, Beer, and Bradley
[1995].

Extending back to AD 1610, the reconstruction is based on a number of
proxy measurements including historical sunspot records and cosmogenic
isotope abundance measurements. The upper panel of figure
\ref{leancompplot} shows this mean annual solar irradiance as a
function of time. Several interesting features are evident in the
reconstruction-- most notably the period of depressed solar output
from 1650-1710, the Maunder minimum. The reconstruction also exhibits
an upward trend, particularly in the twentieth century.


\subsection{Climate Activity}

We adopt as our indicators of climate the multi-proxy surface air
temperature (SAT) reconstructions of Mann {\it et al.}, and Jones {\it
et al.} [Mann, Bradley, and Hughes 1998; Jones, {\it et al.},
1998]. We utilize these temperature reconstructions (rather than using
other climate indicators such as annual precipitation or cloud cover)
because temperature is by far the most extensive and reliably
reconstructed global data set. Using a network of widely distributed
proxy climate indicators, Mann {\it et al.}, and Jones {\it et al.},
have reconstructed the mean Northern Hemisphere surface temperatures
over a time period extending back to AD 1000. The global network of
indicators includes dendroclimatic, ice core, ice melt, coral, and
extensive instrumental records. Reconstructions from individual sites
are spatially averaged to obtain a Northern Hemisphere mean. Using
global reconstructions attenuates the significant overprint by ocean
and atmospheric circulation expected in regional climate records on
the timescales of interest. Figure \ref{mannjones} shows the Northern
hemisphere surface temperature reconstructions of Mann, {\it et al.},
and Jones, {\it et al.} (The two time series reconstructions are not
identical; however, differences in phase and amplitude may be caused
by seasonal and spatial sampling in the reconstructions [Mann, {\it et
al.} 2001]). Both plots are characterized by a gentle cooling that
ends abruptly with a dramatic 20th century warming trend.

\begin{figure}[htbp]
\begin{center}
\includegraphics[scale=1]{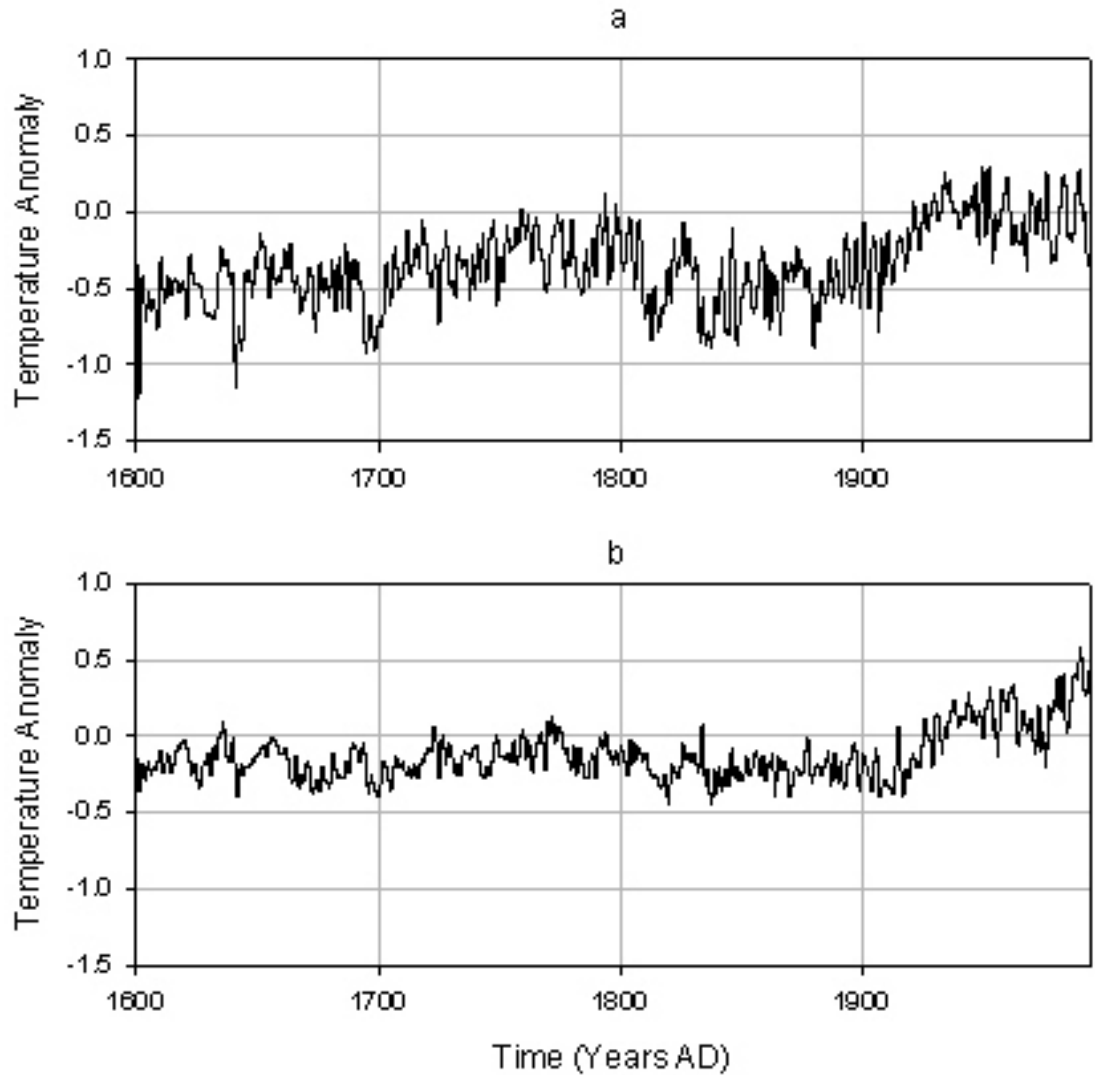}
\end{center}
\caption{Northern hemisphere surface air temperature anomaly (in
$^\circ$C) reconstructions of a) Jones, {\it et al.} [1998], and b)
Mann,
Bradley, and Hughes [1998].}
\label{mannjones}
\end{figure}

\section{Wavelet Analysis}

The basic goal of wavelet analysis is to both determine the frequency
content of a time series, and assess how that frequency content
changes in time. This stands in contrast to conventional Fourier
analysis that allows one to determine what frequencies appear in the
signal, but fails to efficiently reveal the time dependence of those
spectral features. When signals are fraught with localized, high
frequency events, or when phenomena exist on diverse timescales and at
different times, the wavelet transform is the tool of choice. In
recent years, wavelets have found an increasing number of applications
in geophysical and astrophysical research. For example, Weinberg,
Drayson, and Freese have used wavelet analysis to examine the
formation and development of abnormally low total ozone events [1996].
Others have used wavelets to specifically examine solar variability.
Willson and Mordinov have used wavelets to establish sunpots and
faculae as primary causes of total solar irradiance variability
[Willson and Mardinov, 1999]. Continuous wavelet transforms have also
been used by Fligge, Solanki, and Beer to assess the time dependence
of the solar cycle length [Fligge, Solanki, and Beer, 1999], by Frick,
{\it et al.} to analyze stellar chromospheric activity variations
[Frick, {\it et al.}, 1997], and by Weng and Lau to determine
variability of satellite radiance measurements [Weng and Lau, 1994].

\subsection{Basic Theory}

One can expand a complex time signal as the sum of certain basis
functions-- in the case of Fourier analysis, these basis functions are
sines and cosines of different frequencies. Because these functions
are perfectly localized in frequency, it is straightforward to
determine what frequencies are present in the original signal; on the
other hand, because the functions are not at all localized in time, it
is quite difficult to find where in time those frequencies happen to
be. Wavelet analysis circumvents this problem by expanding a time
signal in terms of a set of basis functions localized in both time and
frequency domains. Each of these ``wavelets" is a function of two
parameters: the dilation parameter, $a$, which determines the scale or
frequency of the wavelet function; and the translation parameter, $b$,
which determines the time at which the function is centered. A
wavelet family is generated from a single analyzing, or mother wavelet
by these translations and dilations,

\begin{equation}
{\psi_{ab}=\frac{1}{\sqrt{a}} \psi \left( \frac{t-b}{a} \right) }
\end{equation}

The function $\psi(t)$ must fulfill several conditions in order to be
considered a wavelet. First, it must satisfy the admissibility, or
zero mean condition,

\begin{equation}
\label{admiss}
{\int^{+\infty}_{-\infty} \psi(t) dt = 0}
\end{equation}

Secondly, its localization in both time and frequency domains must
satisfy the following criterion: its spread in time, $\delta t$, and
frequency, $\delta \omega$, must satisfy an uncertainty principle,
$\delta t \delta\omega \ge \mbox{const.}$

The wavelet transform $w_{ab}$ of a signal $f(t)$ is a convolution of
the signal with the wavelet basis functions, and hence depends on the
translation and dilation parameters,

\begin{equation}
{w_{ab} = \int^{\infty}_{-\infty} f(t) \psi \left(\frac{t-b}{a}\right) dt
}
\end{equation}

\noindent We often consider the normalized wavelet power spectrum,

\begin{equation}
{P_{ab} = \frac{|w_{ab}|^2}{\sigma^2}},
\end{equation}

\noindent where $\sigma$ is the variance of the original time series.

There are a number of popular choices for the mother wavelet function
that are well suited to probing geophysical signals. In this work, we
use the second derivative of a Gaussian, the so-called ``Mexican-hat"
function,

\begin{equation}
\label{mexican}
{\psi(t) = c(1-t^2)\mbox{e}^{-t^2/2},}
\end{equation}

\noindent where the constant factor, $c = \sqrt{4/3\sqrt{\pi}}$, ensures
that the wavelet function at each scale has unit energy ({\it i.e.} $\int
\psi^2_{ab}(t) dt = 1$).

In the appendix we address two important technical issues: edge
effects and statistical significance. The preceding analysis works
well as long as we are not interested in spectral features near the
boundary of our signal. Near the beginning and end of finite time
series, however, we need to modify the wavelet basis functions to
avoid spurious results. To attenuate these edge effects, our analysis
is performed using the adaptive wavelets discussed in Appendix A.
Using the methods developed by Frick, {\it et al.}, we allow our
wavelet basis functions to ``adapt" their shape based on the presence
or absence of data in the time series [1997].

Furthermore, in order to discern between essential physical features
of the geophysical signals and those background noise processes that
may mimic them, we must have a means of assessing the statistical
significance of our wavelet transform. Our approach to this issue is
detailed in Appendix B.

As a concrete demonstration of the value of wavelets, consider figure
\ref{sincurv}. The figure depicts a signal consisting of two sine
waves of two different frequencies that are distinctly separated in
time. A conventional Fourier analysis would reveal two peaks, one
corresponding to each frequency, but would offer no information as to
where in time those frequencies dominate. The wavelet transform,
shown in the lower portion of the figure, shows not only the two
different frequencies, but also discloses when each frequency
dominates.

\begin{figure}[htbp]
\begin{center}
\includegraphics[scale=1]{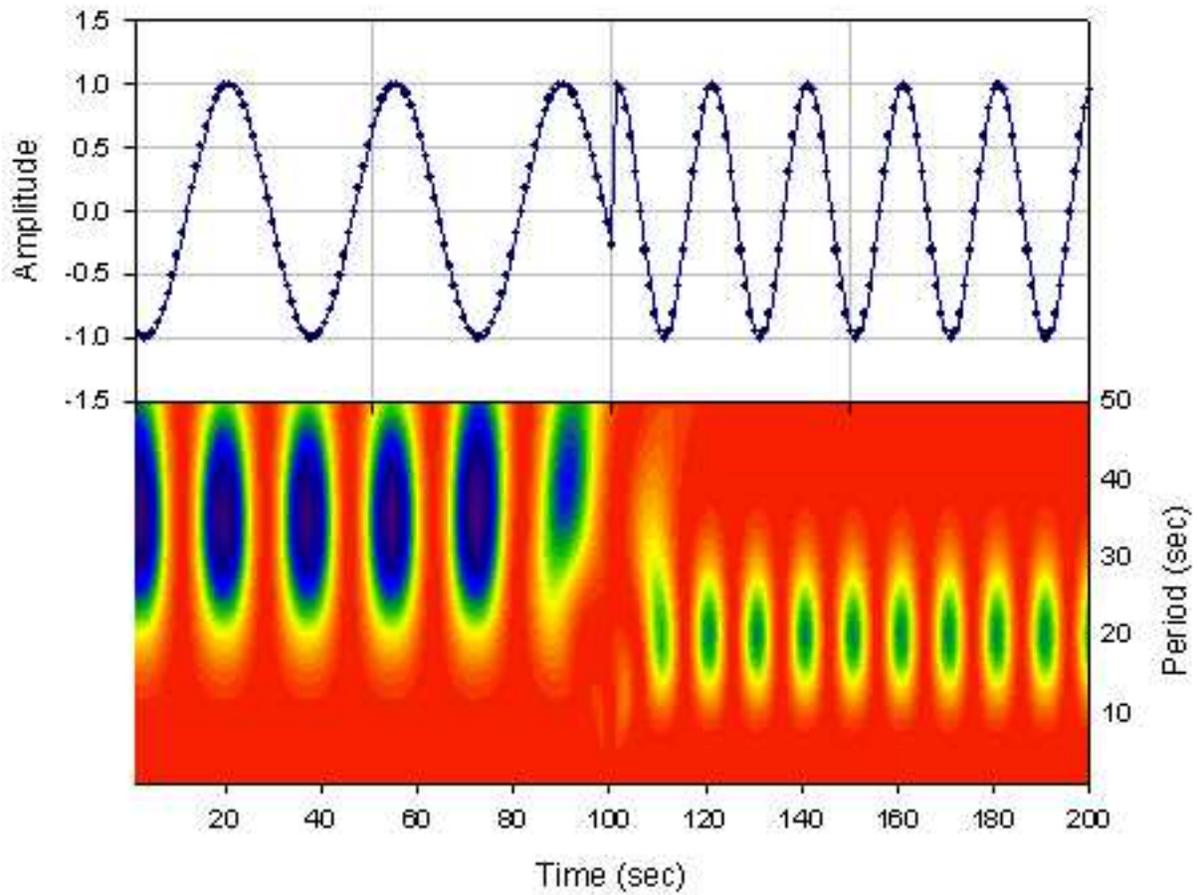}
\end{center}
\caption{Upper panel: a time signal containing two distinct
frequencies. Lower panel: a density plot representing the wavelet
transform of the time signal. From 0 to 100 seconds, the time signal
(upper panel) has a period of 35 seconds, and the wavelet transform
exhibits distinct peaks corresponding to a period of 35 seconds during
this interval. After 100 seconds, the period of the time signal
decreases to 20 seconds and this abrupt change is clearly reflected in
the plot of the wavelet transform. }
\label{sincurv}
\end{figure}

\subsection{Wavelet Transform of Solar Irradiance}

The first step in assessing the sun-climate link is to apply the
wavelet transform to each of the solar and climate reconstructions.
Figure \ref{leancompplot} plots both the solar irradiance
reconstruction of Lean {\it et al.}, and its corresponding wavelet
power spectrum. Significant peaks with a period near 11-years are
evident in the wavelet power spectrum, indicating the presence of the
well-known solar cycle throughout the last several centuries; however,
the solar cycle is notably absent during the Maunder minimum, which
stretches from roughly 1650-1710 [Baliunas, {\it et al.}, 1999].

\begin{figure}[htbp]
\begin{center}
\includegraphics[scale=1]{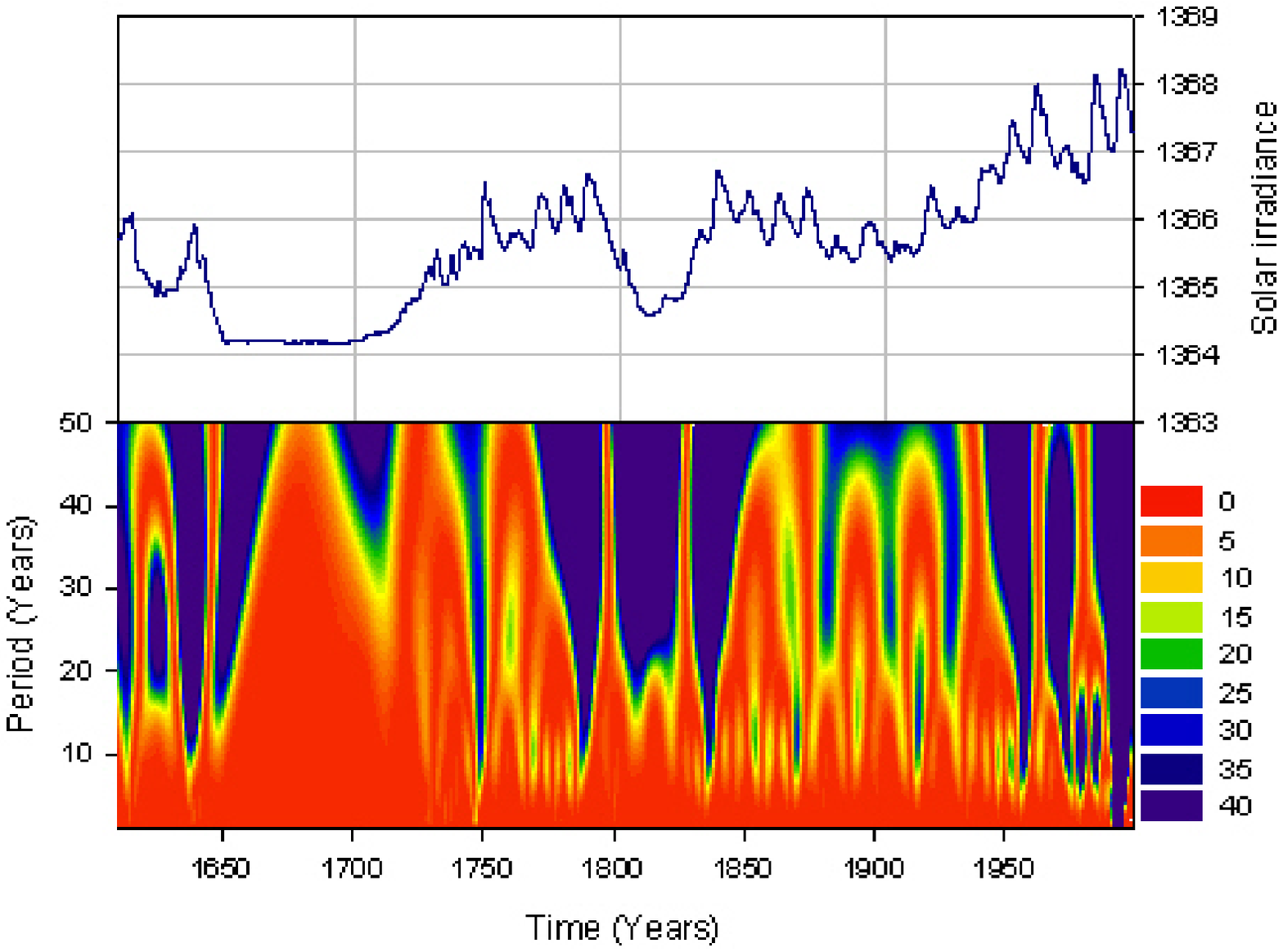}
\end{center}
\caption{Upper panel: the total solar irradiance as reconstructed by
Lean, Beer, and Bradley [1995]. 
Lower panel: the wavelet transform of the solar
irradiance, calculated using the Mexican hat wavelet of
Eq. \ref{mexican}. }
\label{leancompplot}
\end{figure}

Although traditionally considered to be 11-years, the length of the
solar cycle actually changes in time [Ochadlick and Kritikos, 1993;
Fligge, Solanki, and Beer, 1999]. To objectively determine the
frequency content of the solar wavelet transform, we look for peaks in
the wavelet power spectrum $P_{ab}$ corresponding to periods between
three and twenty years. To ensure that the peaks are real, {\it i.e.}
not a result of background noise, we use the statistical significance
tests outlined in Appendix A. We require that the value of $P_{ab}$
at $(a,b)$ exceed the expected red noise background (as given by
Eq. \ref{red noise spectrum}) at the 95\% confidence level. We find
over the past four centuries, that 35 years exhibit significant power
within this range. Following Fligge, {\it et al.}, we chart in figure
\ref{cyclelength} the solar cycle length as a function of time
[Fligge, Solanki, and Beer, 1999]. The mean solar cycle length
obtained by this method is $10.93 \pm 1.53$ years (statistical error).

\begin{figure}[hbtp]
\begin{center}
\includegraphics[scale=1]{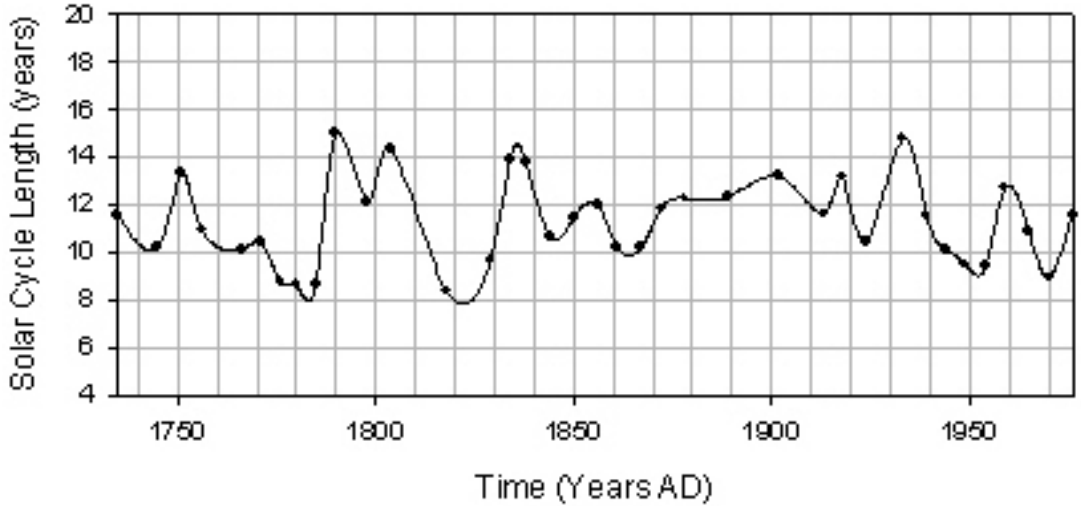}
\end{center}
\caption{Solar cycle length as a function of time, as determined from
the wavelet power spectrum of the solar irradiance reconstruction of
Lean, Beer, Bradley [1995]. }
\label{cyclelength}
\end{figure}

\subsection{Wavelet Transform of Surface Air Temperature}

A similar analysis may be made of the mean annual northern hemisphere
global surface air temperature (SAT) reconstructions. Figure
\ref{manncompplot} plots the SAT reconstruction from Mann, Bradley,
and Hughes [1998] along with its wavelet power spectrum.

\begin{figure}[htbp]
\begin{center}
\includegraphics[scale=1]{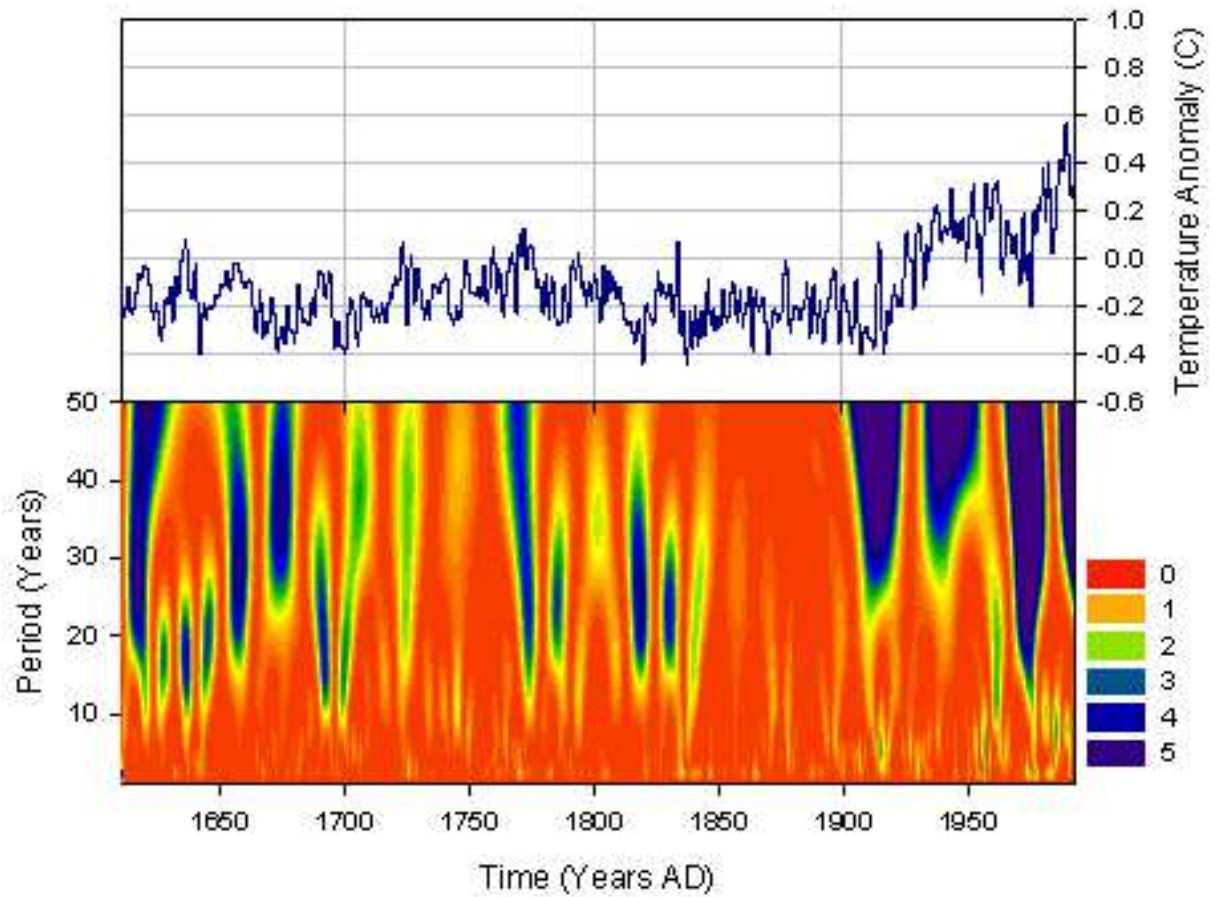}
\end{center}
\caption{Upper Panel: the surface air temperature as reconstructed by
Mann, Bradley, and Hughes [1998]. Lower panel: the wavelet transform of
the
temperature calculated using the Mexican hat wavelet of
Eq. \ref{mexican}. }
\label{manncompplot}
\end{figure}

\section{Wavelet Correlation}

\subsection{Comparison of 11-Year Wavelet Transforms}

We may directly compare the 11-year components of the solar and
temperature wavelet transforms. Figure \ref{elevenlean} plots
horizontal slices through the wavelet transforms of figures
\ref{leancompplot} and \ref{manncompplot} at the 11-year wavelet
period. In the solar irradiance wavelet transform, the periodic
presence of a strong 11-year cycle throughout the past few centuries
is readily apparent, except during the Maunder minimum. On the other
hand, the 11-year cycle is continuously apparent in the Mann
temperature record. The next step in assessing solar-climate forcing
is to quantitatively determine correlations between solar and
temperature wavelet transforms, {\it i.e.} between horizontal slices
such as those depicted in figure \ref{elevenlean}, taken at the
appropriate timescales. Because the length of the solar-cycle varies
somewhat in time, we obtain correlations both for the 11-year
timescales, and also those as determined in figure \ref{cyclelength}.

\begin{figure}
\begin{center}
\includegraphics[scale= 1]{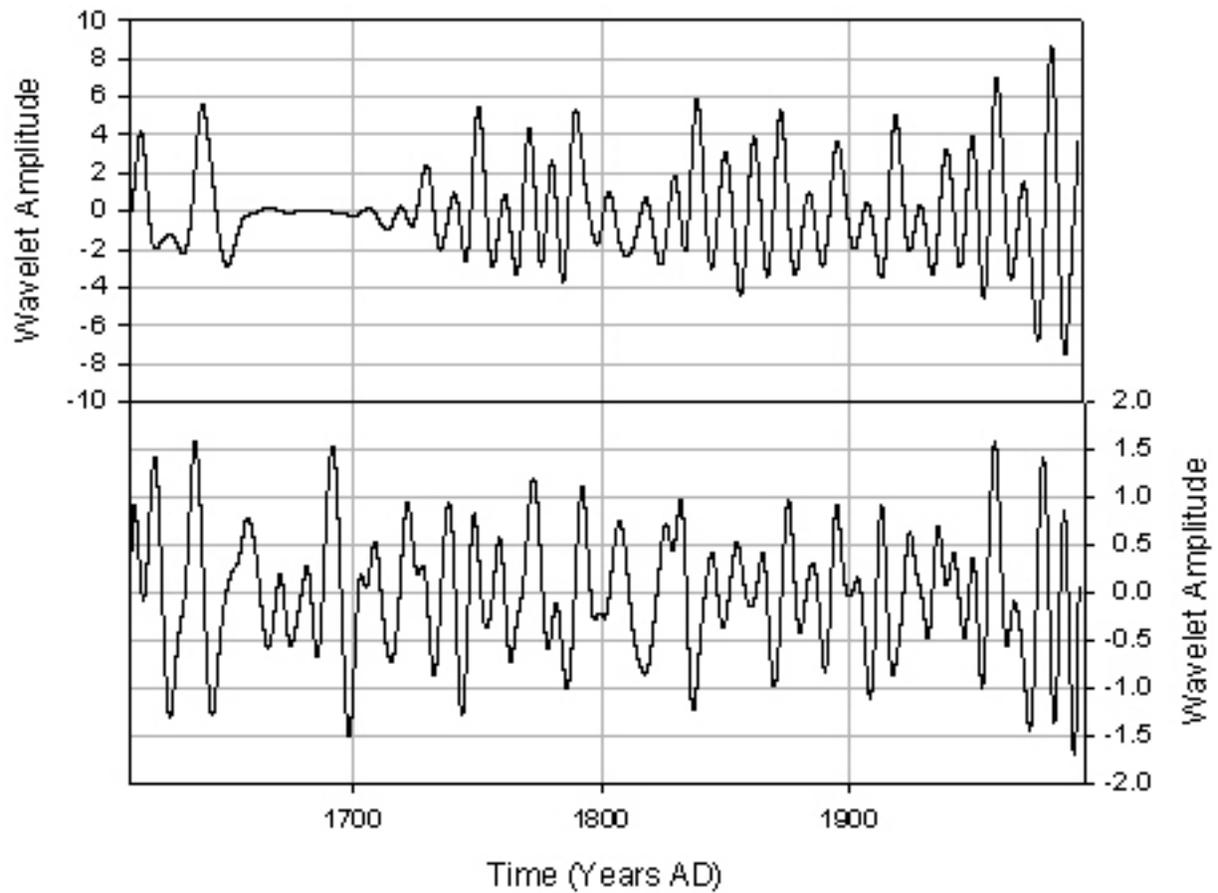}
\end{center}
\caption{Upper panel: the eleven year components of the solar
irradiance wavelet transform. Lower panel: the eleven year components
of the surface air temperature reconstruction of Mann, Bradley, and
Hughes
[1998]. }
\label{elevenlean}
\end{figure}

\subsection{Methods of Correlation}

Once the frequency content of the solar and climate reconstructions
has been established using wavelet analysis, the next step is to
determine how closely the spectral features of one time series
correlate with those of another. To do this we will compare the
spectra using two independent techniques.

\subsubsection{Pearson's Method}

The first of these methods is based on Pearson's correlation
coefficient, which measures the linear correlation between two time
series. That is, for two time series $x(t)$ and $y(t)$, the
correlation between them is given by,

\begin{equation}
{\rho(\phi) = \frac{\sum_t (x(t-\phi) - \bar{x})(y(t) -
\bar{y})}{\sqrt{\sum_t(x(t-\phi)-\bar{x})^2\sum(y(t)-\bar{y})^2} }}
\end{equation}

\noindent where $\bar{x}$ represents the mean of the time series
$x(t)$. Here, $\rho = \pm 1$ indicates a perfect positive or negative
correlation. We have written the correlation as a function of $\phi$,
the phase lag between the two time series. Since terrestrial climate
does not instantly respond to solar fluctuations, the solar irradiance
and temperature records will not necessarily be in phase; in fact,
recent models suggest that surface temperatures can lag irradiance by
up to ten years [Rind, Lean, and Healy, 1999]. By displacing the time
series records with respect to each other by an amount $\phi$, we can
measure the correlation as a function of phase lag.

Furthermore, we have no reason to expect the influence of the sun
on the climate to remain constant in time; in fact, we would like to
be able to track changes in the sun-climate correlation from year to
year. To do this, we measure the correlation between subsets of our
time series defined by a window of width $N$, centered at a time
$t_0$. In the results that follow, we take $N=50$ years. Within the
range $50 \le N/{\rm years} \le 100$ we have verified that the results do not
depend sensitively on the window width; smaller window widths offer
increased time resolution at the expense of statistical significance.
By varying $t_0$, we slide the window along the time series to obtain
a time resolved measure of correlation.

\subsubsection{Conditional Probability Method}

The second correlation method is based on conditional probabilities.
This method has the advantage that is does not depend upon any
assumption of a linear relationship between the two time series.
Given two time series, $x(t)$, and $y(t)$, we consider the following
conditional probability,

\begin{equation}
{P(\phi) = [x(t-\phi)\ge\alpha|y(t)\ge\beta];}
\end{equation}

\noindent that is, the probability that $x(t-\phi)$ is greater than
constant $\alpha$, given that $y(t)$ exceeds some other constant
$\beta$. The constant values $\alpha$ and $\beta$ are typically
chosen to be the mean of the time series under investigation. A large
probability implies that when one signal exhibits a large amplitude,
the other signal does also. Since $P(\phi)$ is a probability, it may
take values between 0 and 1, with zero indicating little correlation,
and 1 indicating a high degree of correlation. This method, as with
Pearson's, may accommodate a phase lag $\phi$; similarly, we can also
achieve time resolution using a sliding window of fixed width. The
conditional probability method requires no assumptions about the
linear relationship between signals, and numerical experiments
indicate that the method is a reliable measure of how closely two time
series track each other.

\section{Results}

We determine the correlation between the wavelet power spectra of the
solar irradiance reconstruction of figure \ref{leancompplot} and the
surface air temperature reconstruction of figure \ref{manncompplot}.
We will begin by comparing the 11-year components of the solar and
temperature wavelet power spectra, which have been plotted in figure
\ref{elevenlean}. First, we will compute correlations for the case
where the irradiance and temperature wavelet spectra are in phase;
this case would correspond to an instantaneous response of the
temperature to changes in the irradiance. After this simplest case,
we will generalize our results in two ways: 1) we will consider the
possibility of a phase lag between solar variation and climate
response, and 2) because the solar cycle length is not constant in
time, we compute the correlation between wavelet components of
different periods. In particular, we will find the correlation for
the appropriate solar cycle length, as plotted in figure
\ref{cyclelength}. For simplicity of presentation, we will focus in
this section primarily on a comparison of Lean solar irradiance and
Mann temperature reconstructions. However, we have in fact performed
identical calculations using the Jones temperature reconstruction as
well; the results will be compared at the end.

\subsection{Comparison of 11-Year Wavelet Power with Zero Phase Shift}

To quantitatively compare the 11-year components of the two wavelet
spectra shown in figure \ref{elevenlean}, we use the Pearson's method
to measure the correlation as a function of time. In figure
\ref{phizeroeleven}, the correlation is plotted as a function of time
with zero phase lag, {\it i.e.}, $\phi=0$. The dashed lines indicate
the 95\% confidence level for the correlation; that is, the
correlation is only significant outside of these lines. We see that,
for most of the data set, the correlation is statistically
insignificant. Additionally, the correlation is both positive and
negative. A positive correlation indicates that an increase in power
in the irradiance wavelet spectrum corresponds to an increase in power
in the temperature wavelet spectrum. A negative correlation indicates
that an increase in irradiance wavelet power is met with a decrease in
temperature wavelet power.

\begin{figure}
\begin{center}
\includegraphics[scale = 1]{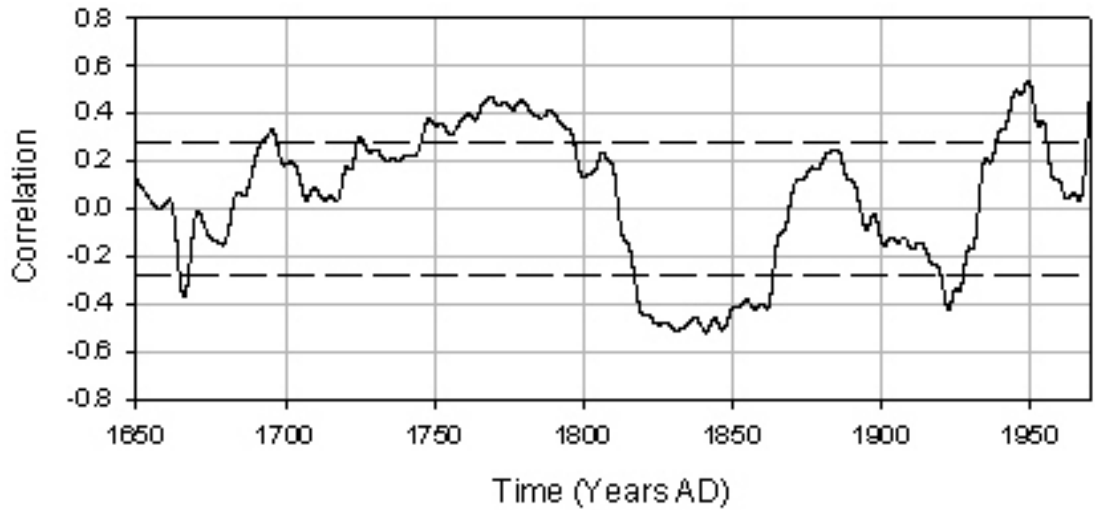}
\end{center}
\caption{The correlation between the 11-year components of solar
irradiance and temperature (Mann {\it et al.} data) wavelet spectra as
a function of time (measured with respect to the solar irradiance
record), for zero phase shift between the signals. The dashed lines
indicate the 95\% confidence level for the correlation; that is, the
correlation is only significant outside of these lines. For most of
the data set, the correlation is statistically insignificant.
Correlation has been obtained using Pearson's method.
}
\label{phizeroeleven}
\end{figure}

However, we do not believe this plot is an accurate characterization
of sun-climate correlation for several reasons. First, we do not
expect the earth's climate to respond instantaneously to solar
variations. There undoubtedly exists a phase lag between irradiance
and temperature; furthermore, the phase lag may vary in time.
Allowing for phase lags radically changes the results. This first
modification is the most important one. Second, although of less
significance, is our exclusive use of the 11-year components in the
above plot. In fact, we know that the length of the solar cycle varies
in time. We can improve our assessment of solar climate forcing by
using the appropriate solar cycle length, as calculated above and
plotted in figure \ref{cyclelength}.

\subsection{Comparison of 11-Year Wavelet Power with Optimal Phase
Shift}\

To demonstrate the sensitivity of our results to the more important of
the two effects, the phase lag, we compute the correlation between
irradiance and temperature wavelet spectra as a function of both phase
lag and time, still for the 11-year components. The result is plotted
in figure \ref{phasecorrelate}.

\begin{figure}[htbp]
\begin{center}
\includegraphics[scale=1]{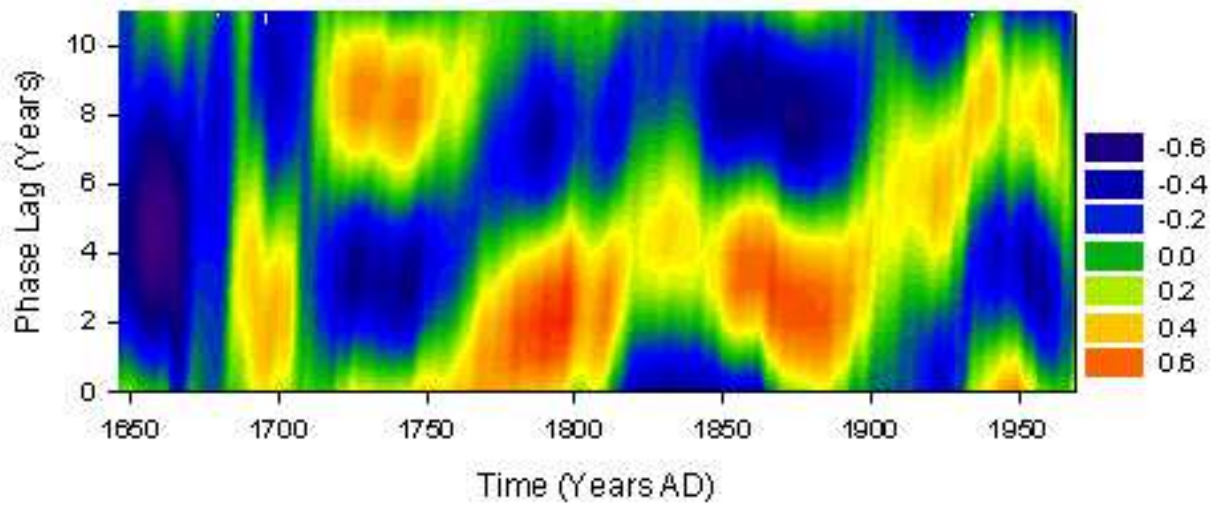}
\end{center}
\caption{The correlation between wavelet power spectra of the
irradiance reconstruction and the surface air temperature
reconstruction (Mann {\it et al.}) at the 11-year wavelet period. The
correlation is plotted as a function of both phase shift and time.
The strength of the correlation is color coded in the range (-1,1),
from strong negative to strong positive correlation. Time is
plotted for the solar irradiance wavelet spectrum, and phase
lag is relative to this time.
Correlation has been obtained using Pearson's method.}
\label{phasecorrelate}
\end{figure}

Figure \ref{phizeroeleven} is the horizontal slice at the bottom of
figure \ref{phasecorrelate}. Whereas the correlation in figure
\ref{phizeroeleven} becomes negative over several intervals, figure
\ref{phasecorrelate} demonstrates that one can always find a positive
correlation by choosing an appropriate phase shift. For each year,
one can find the phase shift that yields the maximum correlation; this
amounts to following the yellow-red ridge in the plot. Because we do
not {\it a priori} know the correct phase shift between the sun and the
climate in a given year, we adopt as an ``optimal" phase shift that
shift that yields the maximum correlation. We require the optimal
phase shift to lie in the range $0 \le \phi \le 10 $ years; we choose
$\phi\ge 0$ because a positive phase shift corresponds to terrestrial
temperature changes lagging behind solar irradiance variations, rather
than the other way around. For the remainder of this paper, we will
always use this optimal phase shift.

Figure \ref{mannlean11phase} plots this maximum correlation as a
function of time, still for the 11-year components. Again, the
dashed-line indicates the 95\% confidence level above which the
correlation is statistically significant. No statistically
significant negative correlations exist now, but strong, significant
positive correlations clearly stand out. In fact, the correlations
are statistically significant throughout most of the range 1720-1955,
{\it i.e.}, after the Maunder minimum and until the recent past.

\begin{figure}
\begin{center}
\includegraphics[scale=1]{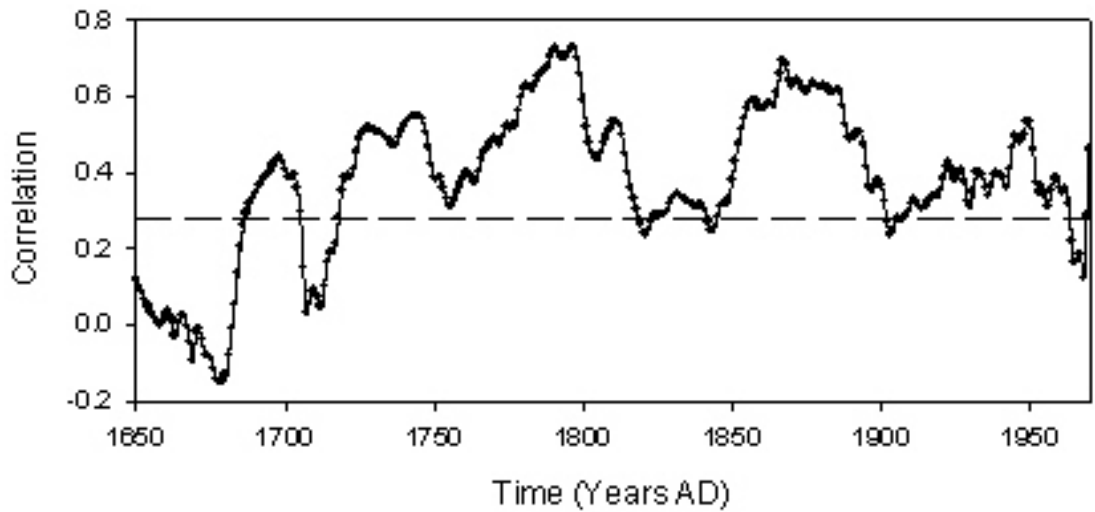}
\end{center}
\caption{The maximum correlation between wavelet power spectra of the
irradiance reconstruction and the surface air temperature
reconstruction (Mann {\it et al.}) at the 11-year wavelet
period. Here, the phase lag between solar output and terrestrial
response has been allowed to vary so as to maximize the correlation.
Time is
plotted appropriate to the solar irradiance wavelet spectrum. The
dashed line indicates the 95\% confidence level above which the
correlation is statistically significant.
Correlation is computed using the Pearson's method.
}
\label{mannlean11phase}
\end{figure}

During the period prior to 1700, the lack of significant correlations
at any phase lag may be due to the presence of the Maunder minimum.
We do not examine temperature and irradiance directly, but rather
their wavelet power spectra; during the Maunder minimum, no power near
the
11-year period exists in the solar record.

\subsection{Comparison of wavelet power spectra with appropriate solar
cycle length and optimal phase shift}
\label{phaseshiftsection}

Because the solar cycle length changes in time, we can further improve
our assessment of solar-climate forcing. Instead of using an 11-year
solar cycle, we now use the corrected solar cycle length plotted in
figure \ref{cyclelength}. We compute correlations between the
irradiance and temperature wavelet spectra at the timescale
corresponding to the corrected solar cycle length; we continue to use
the optimal phase shift ({\it i.e.}, the one that maximizes the
correlation). For example, in the year 1798 we observe a solar cycle
length of 12.11 years in the solar irradiance reconstruction. We
calculate the correlation between the 12.11-year components of the
irradiance and temperature wavelet power spectra at the optimal phase
shift, which in 1798 is 3 years. We can make such a calculation for
each of the 35 years for which we obtained a solar cycle length in
figure \ref{cyclelength}. Figure \ref{mannleancorrect} plots both the
maximum correlation and optimal phase shift for each of these years.

In figure \ref{mannleancorrect}, we see the time dependence of the
optimal phase lag throughout the 400 years of the data sets. Note the
large variability of this phase lag, with values ranging from 0 to 10
years. Of course we may wonder about the veracity of very high
(e.g. 10 year) phase lags. Early on in the data, we find a large
phase lag is required; the phase lag dips to be quite small for about
150 years, and around 1900 begins to rise again.

\begin{figure}
\begin{center}
\includegraphics[scale=1]{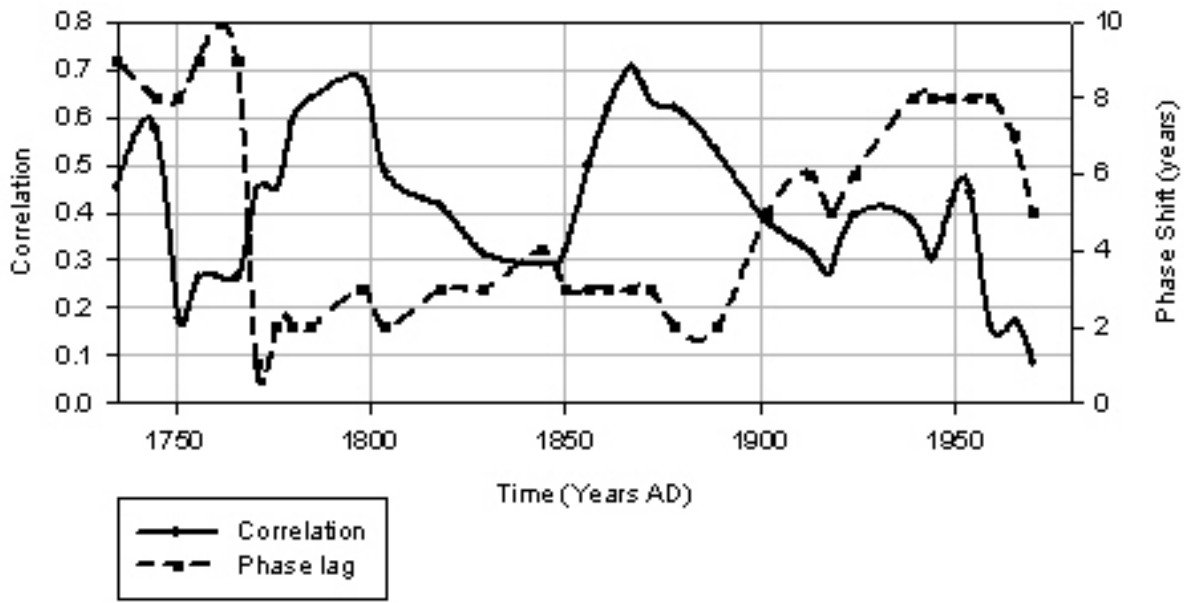}
\end{center}
\caption{The solid line tracks the correlation for each of the 35
years for which we obtained a solar cycle length in figure
\ref{cyclelength}; the solid line shows the
maximized correlation between the
wavelet power spectra of the surface air temperature reconstructions
of Mann {\it et al.} and the solar irradiance reconstruction.
The dotted line tracks the optimal phase shift, {\it i.e.}, the phase
shift necessary to achieve maximum correlation.
Time is appropriate to the solar irradiance spectrum.
Correlation has been computed using the Pearson's method.}
\label{mannleancorrect}
\end{figure}

All of the results we have presented in this section have used only
the Mann {\it et al.} surface air temperature reconstruction, but we
have also performed the same analysis with the Jones {\it et al.} SAT
reconstruction. For comparison, the maximum correlations of both the
Mann {\it et al.} and Jones {\it et al.} are plotted together in
figure \ref{mannjonescorr}.

\begin{figure}
\begin{center}
\includegraphics[scale=1]{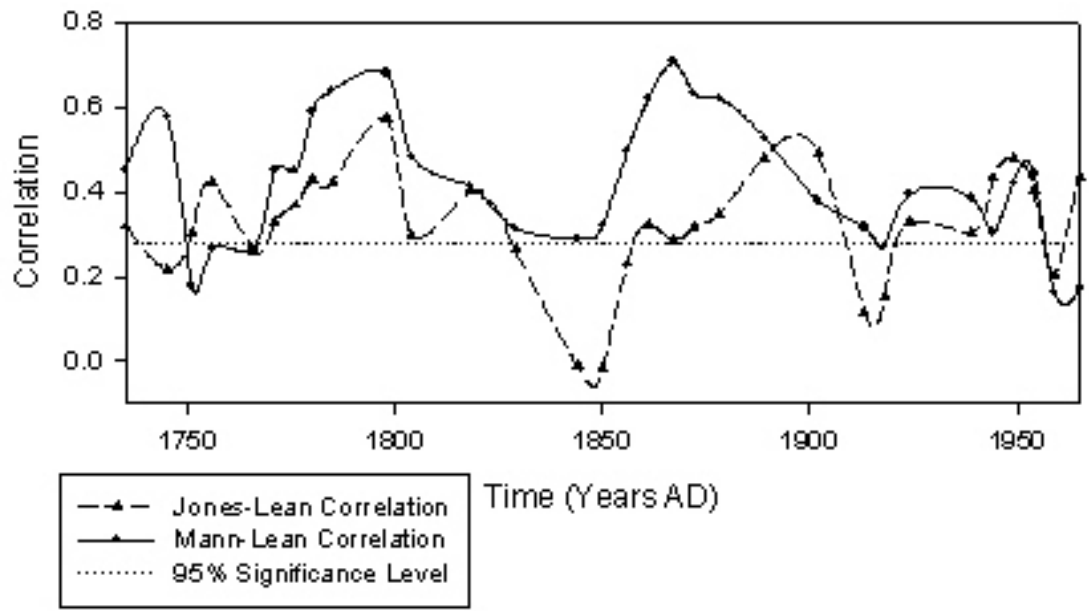}
\end{center}
\caption{The solid line plots the maximum correlation between wavelet
spectra of solar irradiance and the surface air temperature
reconstructions of Mann {\it et al.}. The dashed line plots the same
correlation for the temperature reconstruction of Jones {\it et
al.} The straight dotted line represents the 95\% confidence level
above which correlations are statistically significant.
Time is appropriate to the solar irradiance spectrum.
Correlation has been computed using the Pearson's method.}
\label{mannjonescorr}
\end{figure}

\subsection{Comparison of wavelet spectra for a variety of timescales
(with optimal phase shift)}

The focus of this paper is the solar cycle timescale of roughly 11
years. However, the wavelet transforms that we have obtained to date
of the solar irradiance and temperature data have yielded much
additional information, on scales from 1-50 years. Hence we can
investigate the sun-climate correlation on all of these timescales
simultaneously by plotting the maximum correlation as a function of
time and wavelet period. The results are shown in figure
\ref{MaxCorVsPer}. As before, we allow $ 0\le \phi \le 10 $ years. In
future work we will explore even longer timescales of centuries to
millenia with these wavelet techniques.

\begin{figure}[htbp]
\begin{center}
\includegraphics[scale=1]{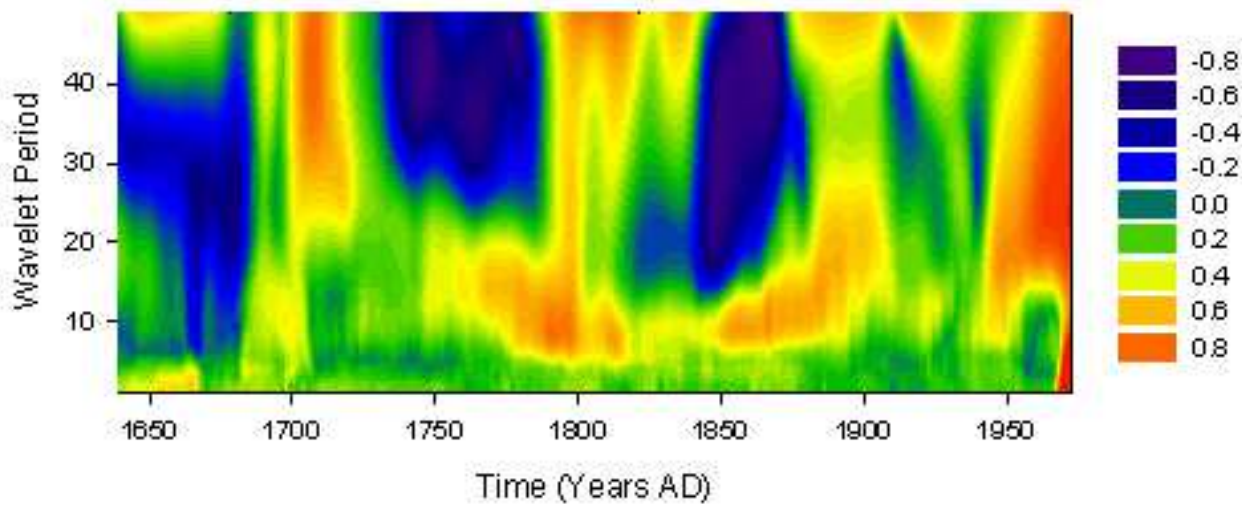}
\end{center}
\caption{Maximum correlation between solar irradiance and temperature
(Mann {\it et al}) wavelet spectra, as a function of both time
(appropriate to the solar irradiance reconstruction) and
wavelet period.
Correlation has been computed using Pearson's method.
}
\label{MaxCorVsPer}
\end{figure}

\subsection{Results using the Probability correlation method}

The preceding analyses have all utilized the Pearson's correlation
method. In addition, we have performed the same calculations with the
probability correlation method, as a check on our results. Figure
\ref{probcorrelation} compares the results of applying Pearson's
correlation method with that obtained by applying the probability
method. In both cases, we compute the correlation between the wavelet
power spectra of the Mann, {\it et al.} SAT reconstruction, and the
Lean, {\it et al.}, solar irradiance reconstruction, adjusted for
optimal phase shift and the appropriate solar cycle length. Although
the two measures of correlation are not identical, both contain
similar features as a function of time. Since the probability
correlation method does not depend on a linear relationship between
the two time series, it is encouraging that both methods give the same
results.

\subsection{Phase Lags}

Several authors have previously addressed the phase relationship
between solar forcing and climate response.  Their results differ
markedly from ours.  We believe our wavelet analysis to be superior to
any previous work on this subject.  Thomson claimed moderate, in-phase
coherence between Northern Hemisphere global temperatures and sunspot
records from 1854 to 1923; from 1923-1991 the same records suggest
temperatures 180$^\circ$ out of phase with the solar cycle [Thomson,
1995]. Analyzing the same data, Lawrence and Ruzmaikin reported very
different results: they claimed a negative coupling with a phase shift
of $\phi \approx -135^\circ$ ({\it i.e.,} with temperature changes
leading solar changes) during the period 1860-1920, switching abruptly
to a positive coupling ($\phi \approx 45^\circ$) from 1920 to the
present [Lawrence and Ruzmaikin, 1998]. Furthermore, studies by Currie
indicate a possible geographic dependence of the solar-climate phase
[Currie, 1993]. Examining United States temperature records, Currie
finds that cyclic temperature variations are in phase with solar
irradiance East of the Rocky Mountains, while west of the Rockies they
are 180$^\circ$ out of phase.

Our work instead uses wavelet analysis to determine an optimal phase
lag.  This approach has many advantages. By using wavelet transforms
of the data we are able to obtain information exclusively on the
timescales of interest, namely the solar cycle timescale; the
timescale itself is obtained using wavelets (see figure
\ref{cyclelength}).  Using wavelets we are clearly able to identify
the power on solar cycle timescales at different times, and are able
to compare the power in the solar irradiance and terrestrial
temperature data.  The results of our study are plotted in figure
\ref{mannleancorrect}. Roughly, we see that temperature records lag
solar irradiance records by roughly 90$^\circ$ from 1760 to 1878, then
become nearly 180$^\circ$ out of phase until 1924; from 1924-1970, the
phase lag is nearly 270$^\circ$. Note that our study only uses global
mean averages of the annual data; we have not studied geographic
variations.

\begin{figure}[htbp]
\begin{center}
\includegraphics[scale=1]{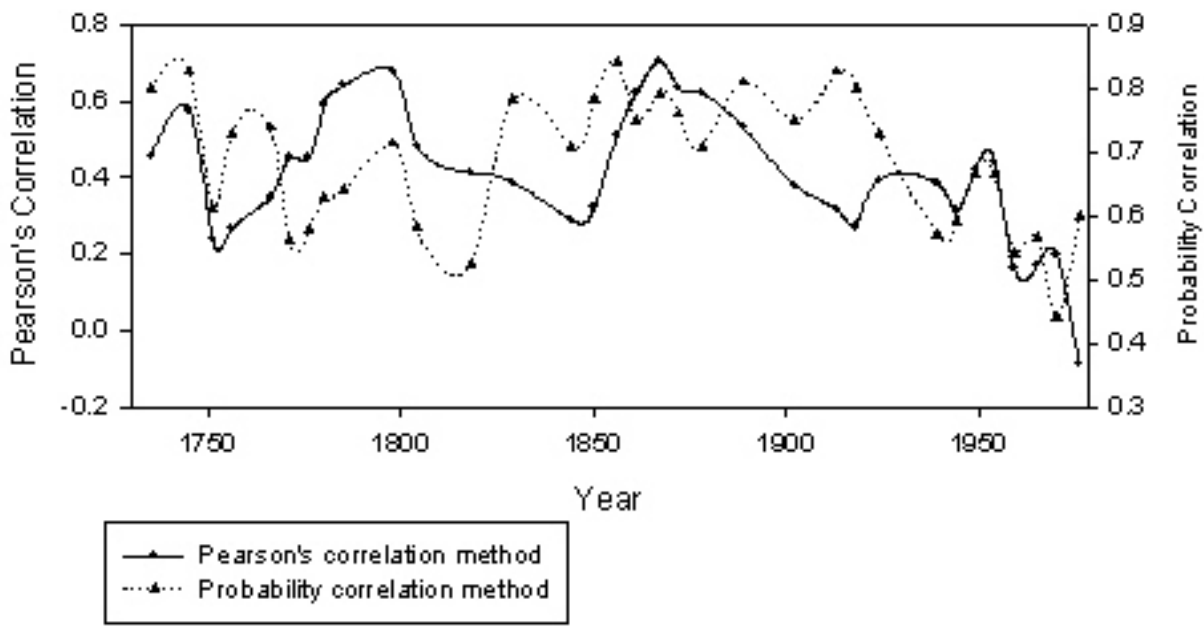}
\end{center}
\caption{The solid line represents the Pearson's correlation between
the wavelet power of the surface air temperature reconstruction of
Mann {\it et al.} and the solar irradiance reconstruction. The
dashed line tracks the correlation between the same power spectra
using the probability correlation method, using the same optimal phase
shifts. Time is plotted appropriate to the solar irradiance
reconstruction. }
\label{probcorrelation}
\end{figure}

\section{Stochastic resonance in solar climate forcing: speculation}

The results of the previous sections reveal the presence of a time
dependent correlation between the terrestrial temperature and solar
irradiance, but do not provide a mechanism by which this sun-climate
correlation may vary. In fact, the means by which apparently
insignificant variations in the incident solar radiation seem to
effect relatively large changes in the Earth's temperature remains an
outstanding problem of climate science. The phenomenon of stochastic
resonance has been proposed as a mechanism for enhancing the effects
of weak forcings[Benzi, {\it et al.}, 1982]. In particular, a weak
periodic forcing signal is amplified by the noise associated with a
nonlinear system [Gammaitoni, {\it et al.}, 1998; McNamara and
Wiesenfeld, 1989; Bulsara and Gammaitoni, 1996]. Lawrence and
Ruzmaikin have recently proposed that the phenomenon of stochastic
resonance may be an important factor in the solar forcing of the
climate on 11-year timescales [Lawrence and Ruzmaikin, 1998]. We here
illustrate a method (using wavelet techniques) that allows us to test
whether or not stochastic resonance has helped to drive solar climate
forcing.

Stochastic resonance is a mechanism whereby a weak, periodic forcing
signal is amplified by the noise associated with a nonlinear system
with more than one minimum. The classic example of stochastic
resonance involves an overdamped particle moving in a noisy, quartic
potential well,

\begin{equation}
{U(x) = -\frac{a}{2}x^2 + \frac{b}{4}x^4}
\end{equation}

\noindent where $a$ and $b$ are constants that depend on the nature of
the problem. The minima for the potential are located at $x = \pm
\sqrt{a/b}$, and the height of the potential barrier is $a^2/4b$.

\begin{figure}
\begin{center}
\includegraphics[scale=1]{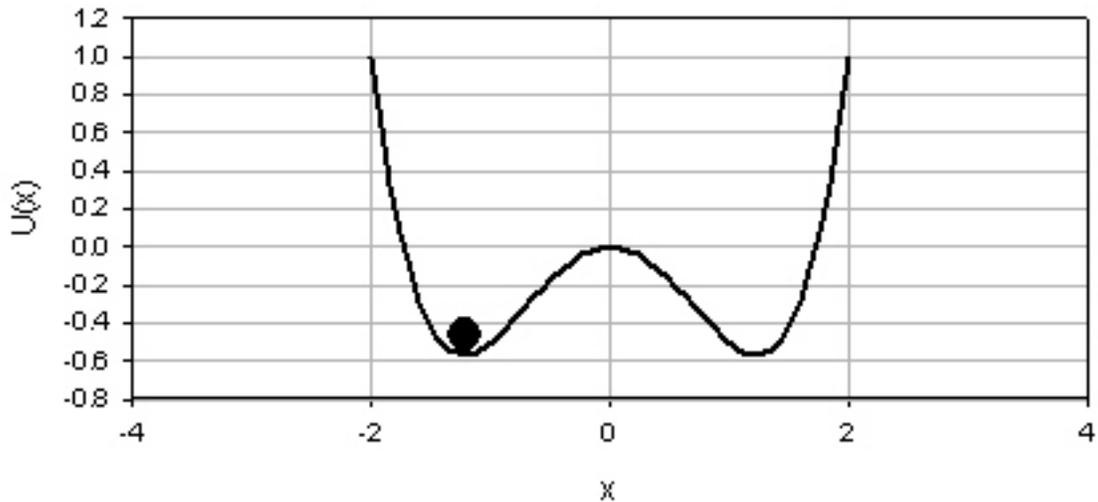}
\end{center}
\caption{A bistable, quartic potential.}
\label{bistable}
\end{figure}

The potential is plotted in figure \ref{bistable} for a=1.5 and b=1.
The particle naturally resides in one of the two stable minima, but an
external forcing signal may provide the energy necessary for the
particle to move over the potential barrier and down into the adjacent
minima. In a noiseless system, if the external forcing signal is too
weak, the particle will always remain in one of the minima; however,
in a noisy system, the particle may exploit some of the noise energy
to leap from minimum to minimum, in phase with the weak forcing
signal. In a sense, the weak signal cooperates with the noise to
induce a relatively large change in the system.

Lawrence and Ruzmaikin propose that under certain conditions,
terrestrial climate systems exhibit a bistability that allows
stochastic resonance to amplify a weak 11-year solar signal, thereby
producing ``transient correlations" between the sun and the
terrestrial climate. Whether an appropriate bistability exists in the
climate system is unclear, but Ruzmaikin suggested that the ENSO
oscillations may act as a stochastic driver, exciting transitions
between normal atmospheric states and anomalous ones such as the
Pacific North American (PNA) pattern [Ruzmaikin, 1999].

To test whether the observed, transient correlations of figure
\ref{mannleancorrect} could be induced by stochastic resonance
(regardless of the physical nature of the bistability), we look for a
``hallmark" feature of stochastic resonance: the response of the
system to a driving signal depends on the intensity of the noise in
the system. It is a general feature of stochastic resonance models
that the response of the system peaks at a particular value of the
noise intensity.

For a bistable system, such as the quartic potential discussed
above, one can show that, under certain circumstances, the dependence
of response on noise at the fundamental forcing frequency is,

\begin{equation}
{\mbox{response} \propto
\left(\frac{Ac}{D}\right)^2\exp(-U_{\mbox{max}}/D)}
\end{equation}

\noindent where $A$ is the amplitude of the driving signal, $c$ is the
separation of the minima, $U_{\mbox{max}}$ is the height of the
potential barrier, and D is the intensity of the noise in the system.
This dependence is plotted in figure \ref{stochastictheory}. Although
details of the signal response curve depend on the specific system,
the fact that response varies with noise intensity is a ``fingerprint"
feature of stochastic resonance models.

\begin{figure}
\begin{center}
\includegraphics[scale=1]{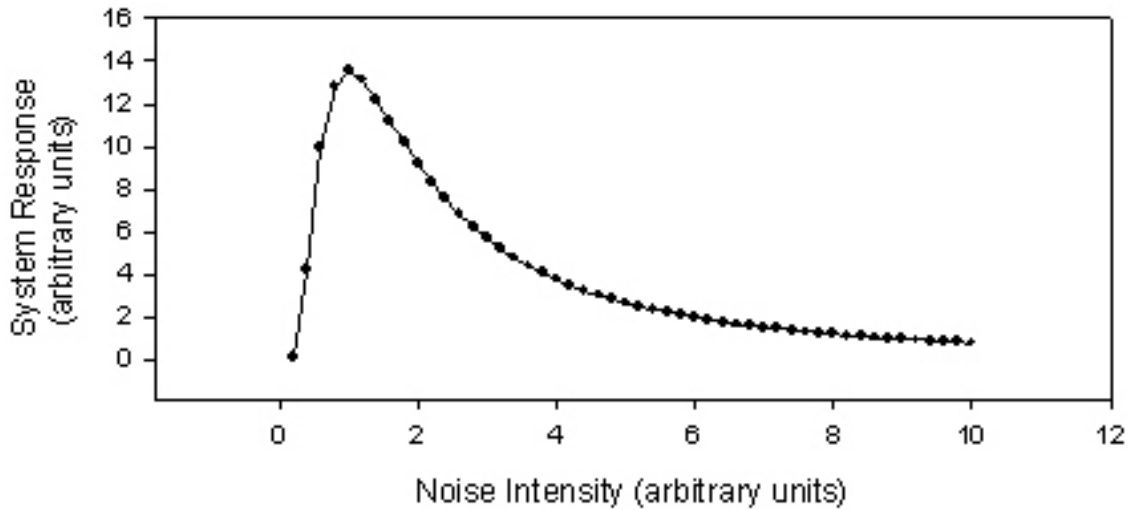}
\end{center}
\caption{The response, as a function of noise intensity, of a bistable
nonlinear system driven with a weak, periodic signal. The peak
indicates stochastic resonance.}
\label{stochastictheory}
\end{figure}

To test whether or not the solar climate system exhibits stochastic
resonance, we generate an analogous plot from our data and compare it
with figure \ref{stochastictheory}. We take the sun-climate
correlation (between wavelet power spectra on ``the adjusted 11-year"
time scale), as plotted in figure \ref{mannleancorrect}, as a measure
of system response. Our system thus exhibits two states: one
corresponding to a high sun-climate correlation, and the other to a
small sun-climate correlation. As a crude measure of noise intensity,
we use the variance of the original temperature time series
reconstruction. In other words, we use the variations in Earth
temperature as a measure of the noise inherent to the climate system.
We will investigate whether these changes in temperature variance
strongly affect the magnitude of the sun-climate correlation.

Because the variance of the temperature record changes in time, we can
track its effect on the sun-climate correlation. To obtain a plot
analogous to figure \ref{stochastictheory}, we plot the sun-climate
correlation in a fifty-year window versus the noise in that window.
The result is shown in figure \ref{stochasticproper}.

\begin{figure}[ht]
\begin{center}
\includegraphics[scale=1]{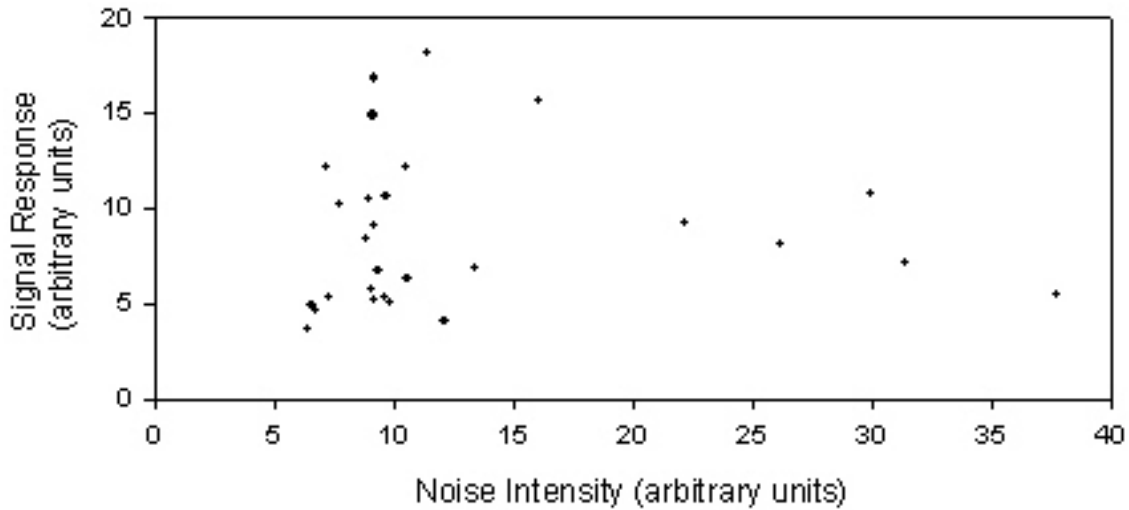}
\end{center}
\caption{The sun climate correlation as a function of the noise
inherent to the system, as estimated by the variance of terrestrial
temperature.}
\label{stochasticproper}
\end{figure}

The plot suggests that the sun-climate correlation varies as a
function of noise intensity, and it even bears a passing resemblance
to figure \ref{stochastictheory}. However, we must be cautious: the
range of temperature variance is quite small, and most of the points
on the plot have a variance between 5 and 15 units. Unfortunately, we
are unable to tune the noise as one would in an experiment. We have
only a limited terrestrial temperature record to work with. Given
this data set, we have no ability to change the intensity of the noise
background; we are limited by the natural nonstationarity inherent to
the system. Furthermore, we have no reason to believe that a simple
bistable potential (such as the one that gives rise to figure
\ref{stochastictheory}) adequately represents the terrestrial climate.

There are a number of additional caveats. The use of the variance of
the temperature reconstruction as a proxy for noise intensity is most
likely naive. The relevant stochastic background may not be noise
fluctuations associated with the temperature itself, but rather with
some other stochastic phenomenon, such as the ENSO driver, as
suggested by Ruzmaikin. Furthermore, assuming that the sun-climate
correlation is a good indicator of system response may also be
inadequate. Detailed climate models sensitive to stochastic resonance
type phenomena may help assess the validity of these ideas.

The transient sun-climate correlation depicted in figures
\ref{mannlean11phase} and \ref{mannleancorrect}, exhibits a period on
the order of 50 years. Because we are examining the sun-climate
correlation on 11-year time scales, the presence of such a long period
is difficult to explain. Systems exhibiting stochastic resonance are
most likely to switch states when the periodic driver peaks; thus, in
our case, we would expect the sun-climate correlation to switch from
high to low (or {\it vice-versa}) on average every, $11/2 \approx 5.5$
years. Longer term solar cycles (such as the 88-yr Gleissberg cycle)
may force the climate system to switch states on 40-50 year scales,
producing the observed transient correlations. Alternatively, because
the 11-year forcing signal is so weak, the earth may reside in one
state for several ({\it i.e.} 4-5) forcing periods before
transitioning to the other. In future work we will use wavelets to
examine other tests of stochastic resonance in the Earth/Sun system.

Hence our analysis does not exclude stochastic resonance as a possible
mechanism by which solar-climate forcing proceeds; in fact, the
presence of transient sun-climate correlations and a possible
dependence of the sun-climate correlation on background noise
intensity suggests that stochastic resonance may be present.

\section{Conclusion}

We have used wavelet analysis as a tool to investigate the response of
terrestrial temperatures to solar variations. Wavelet transforms of
solar irradiance and surface air temperature reconstructions are
plotted in figures \ref{leancompplot} and \ref{manncompplot}, and
reveal the time dependence of the spectral content in those signals. From 
the wavelet transform of the solar irradiance reconstruction, we
objectively determine the solar cycle length as a function of time by
looking for peaks in the wavelet spectrum. Using two different
methods, we examine the correlation between the wavelet power spectra
of the irradiance and temperature records. In this paper we have
focused on solar-cycle timescales (near 11 years). In future work, we
will extend the analysis to include centennial and millennial-scale
variations.

Determining the correlation between the spectral content of two
signals is more robust than comparing directly the raw time signals:
the wavelet transform acts as a filter, allowing us to examine
frequencies of interest while excluding much of the stochastic
variability and background noise inherent in the original signal. In
addition, we can assess the strength of the correlation on individual
timescales, and track that correlation as a function of
time.

We have computed the correlation as a function of both phase $\phi$
and time t, where $\phi$ is the phase shift between the solar
irradiance and terrestrial temperature wavelet power spectra ({\it
i.e.}, in some sense, the phase lag between solar forcing and climate
response).

\subsection{Summary of Results}

First, we studied the 11-year components of the solar irradiance and
surface air temperature wavelet power spectra. We found that, for
$\phi=0$ (no phase shift between the signals), there is almost no
statistically significant correlation between these components, as
shown in figure \ref{phizeroeleven}. This result is not surprising,
because we expect that the earth takes some time to respond to changes
in the sun. A time dependent phase shift must be introduced to obtain
a positive correlation of any significance.

Next, we allowed for phase shifts between the solar irradiance and
terrestrial temperature wavelet spectra. We looked for the optimal
phase shift, {\it i.e.}, that phase shift that maximizes the
correlation. We plotted the correlation as a function of phase lag and
time for the 11-year components in figure \ref{phasecorrelate}. By
choosing an appropriate phase for each year (following the
yellow-red ridge in figure \ref{phasecorrelate}), we observe that it
is possible to achieve a positive correlation. Indeed, we found a
significant, positive sun-climate correlation for most of the period
AD 1720-1950 ({\it i.e.} after the Maunder minimum and until the
recent past). This result is plotted for the 11-year timescale in
figure \ref{mannlean11phase}.

Because the solar cycle length varies in time, we improved our results
further by finding the appropriate solar-cycle length, as plotted in
figure \ref{cyclelength}, and computing the irradiance-temperature
wavelet spectra correlation at that timescale. This correlation is
presented in figures \ref{mannleancorrect} and
\ref{mannjonescorr}. One can see that the phase-optimized correlations
of figures \ref{mannlean11phase}, \ref{mannleancorrect}, and
\ref{mannjonescorr} oscillate in time.

In the case of the sun-climate correlation corrected for phase lag and
solar cycle length, shown in figure \ref{mannleancorrect}, we found
that the strength of the correlation is not at all constant in time:
in fact, the correlation increases and decreases between 0.12 and 0.71
over several intervals throughout the past 400 years. One could
speculate that there exists some periodicity in the strength of the
earth's response to solar irradiance. The time dependence of the phase
lag between earth and sun that would give the maximum correlation
between the two signals is plotted over the 400 years of the data set
in Figure 13.

Additionally, we plotted the correlation between the irradiance and
temperature for wavelet period from 1-50 years, for the optimal phase
shift in the range $0<\phi<10$ years. We also speculated on the role
of stochastic resonance in this variability in the sun-climate system
and presented a possible test for this effect. }

\appendix

\section{Wavelet Analysis: Edge Effects}

{ When we are interested in spectral features near the boundary of our
time series, we must consider edge effects. Near the beginning and
end of a finite time series, the wavelet function becomes unable to
satisfy the admissibility condition of Eq. \ref{admiss}, leading to
spurious modifications of the wavelet transform in the vicinity of the
boundaries. To attenuate these edge effects, our analysis is
performed using adaptive wavelets. Using the methods developed by
Frick, {\it et al.}, we allow our wavelet basis functions to ``adapt"
their shape based on the presence or absence of data in the time
series [1997].

We consider a time series, $f(t)$, that is defined over some finite
time interval, and which may contain a finite number of ``gaps", where
no data has been recorded. To keep track of observed data, we define
a gap function, $G(t)$, that is equal to one everywhere data is
present and zero otherwise. Using this approach, the boundaries are
thus viewed as semi-infinite gaps in the signal. We replace our
wavelet $\psi(t)$ with a modified analyzing wavelet, $\psi^\prime
(t)$.

\begin{equation} {\psi^\prime(\frac{t-b}{a}) = \psi(\frac{t-b}{a})G(t)}
\end{equation}

Near a gap, this modified wavelet ``breaks", becoming unable to
satisfy the admissibility condition of Eq. \ref{admiss}. In this
manner we shift the blame from the incomplete data set $f(t)$, to an
improperly constructed wavelet function $\psi^\prime(t)$.

To fix the problem we replace $\psi^\prime$ with an {\em adaptive}
wavelet, $\tilde{\psi}$, which changes shape to guarantee that the
admissibility condition is always satisfied. We accomplish this by
writing our analyzing wavelet as the product of two functions,

\begin{equation} {\psi = h(t)\Phi(t)} .
\end{equation}

In the case of the Mexican hat wavelet used in this analysis,
$\Phi(t)$ is the positive definite Gaussian function, exp$(-t^{2}/2),$
and $h(t)$ represents the modulating parabola $c(1-2t^{2})$. The
adaptive wavelet is then defined as,

\begin{equation} \tilde{\psi}(t,b,a)=\left[ h\left( \frac{t-b}{a}\right)
-C(a,t)\right] \Phi\left( \frac{t-b}{a}\right) G(t)\label{Adapted
Wavelet} .
\end{equation}

The function $C(a,t)$ is determined by requiring that $\int\tilde{\psi
}(t)dt=0$

\begin{equation} C(a,t) =\left[ \int_{-\infty}^{\infty}\Phi\left(
\frac{t-b}{a}\right) G(t)dt\right] ^{-1}\times\nonumber\\
\int_{-\infty}^{\infty}h\left( \frac{t-b}{a}\right) \Phi\left(
\frac{t-b}{a}\right) G(t)dt .
\end{equation}

The gap function must be determined specifically for each data set.
In practice, the reconstructed time series are padded with zeros, with
the gap function specifying the limits of the actual data. This
technique reduces but does not eliminate the error; one must still be
wary of spectral features that fall near the boundaries of the time
series. Our results will not rely on those portions of the wavelet
transform strongly affected by residual edge effects.

\section{Wavelet Analysis: Background noise and statistical significance}

In order to discern between essential physical features of the
geophysical signals and those background noise processes that may
mimic them, we must have a means of assessing the statistical
significance of our wavelet transform. Wavelet statistical
significance tests are accomplished by comparing the normalized
wavelet power spectrum $P_{ab}$ with the expected noise power spectrum
$P_{noise}$. Thus, in order to judge statistical significance, we must
first understand the background noise associated with our signal.
Geophysical noise backgrounds are often modeled as either white noise
(having a flat Fourier spectrum), or red noise (having more power at
lower frequencies).

Following Torrence and Compo, the background noise in the time series
reconstructions used in this paper are modeled as first order
autoregressive (AR-1) processes, also known as a Markov process
[1998]. The simplest model for an AR-1 time series $x_{t}$ is,

\begin{equation} x_{t}=\alpha x_{t-1}+z_{t},\label{markov} \end{equation}

\noindent where $z_{t}$ is normally distributed white noise, and the
parameter $\alpha$ is the lag-1 autocorrelation coefficient of the .

It can be shown that the expected Fourier spectrum for an AR-1 noise
process is given by,

\begin{equation}
P_{AR-1}(k)=\frac{1-\alpha^{2}}{1+\alpha^{2}-2\alpha\cos(2\pi k/N)}
\label{red noise spectrum}
\end{equation}

\noindent where $N$ is the number of points in the time series and $k$
is the frequency index. Because the background contains more power at
lower frequencies, it is termed a \emph{red-noise} spectrum. Numerical
experiments performed by Torrence and Compo bear out the hypothesis
that the spectrum of Eq. (\ref{red noise spectrum}) is also the
expected \emph{local wavelet power spectrum} at a time $b_{0}$ ({\it
i.e.}, the function $P_{noise}(a)=|w_{a,b=b_{0} }|^{2}$) [1998].

Torrence and Compo also confirm, by Monte Carlo simulation, that the
actual noise spectrum is $\chi^{2}$-distributed about the predicted
mean background spectrum. \ In particular, for real wavelets, such as
the Mexican-hat used in this analysis,

\begin{equation} P_{noise}\rightarrow
P_{AR-1}(k)\chi_{1}^{2}\label{expected noise} \end{equation}

\noindent where the arrow indicates ``is distributed as''.

The noise background of our signals can be estimated by determining
the lag-1 autoregressive coefficient, $\alpha$, appearing in
Eq.(\ref{markov}). The coefficient $\alpha$ is estimated from the
original geophysical time, $y_{i}$, series by

\begin{equation}
\alpha=\frac{\sum_{i=1}^{i=N}(y_{i}-\bar{y})(y_{i-1}-\bar{y})}{\sum
_{i=1}^{i=N}(y_{i}-\bar{y})^{2}}\label{autoregression} .
\end{equation}

By assessing the statistical significance of our results, we may
filter those regions of the wavelet spectra associated with stochastic
variability and known noise processes, thereby enhancing our ability
to detect real signals. Knowing the expected power spectrum and the
distribution, we may establish levels of confidence to distinguish a
physical component of the wavelet spectrum from the noise
background. We regard the wavelet spectrum as significant if it
exceeds the expected noise background of Eq. (\ref{expected noise}) at
the 95\% confidence level.

}
\end{document}